\documentclass[12pt]{iopart}
\usepackage{graphicx} 
\usepackage{cite} 
\usepackage{float}
\usepackage{caption}
\usepackage[colorlinks=true,linkcolor=blue,citecolor=blue,urlcolor=blue]{hyperref} 
\begin{document}

\title{Deciphering Noise in tip--sample Interactions: Insights into Nanoscale Dynamics}

\author{Jaime Colchero$^1$, Juan F. González-Martínez$^2$}
\address{$^1$Universidad de Murcia, Departamento de Física Aplicada}
\address{$^2$Universidad Politécnica de Cartagena, Departamento de Física Aplicada y Tecnología Naval}
\ead{colchero@um.es} 

\begin{abstract}

Noise sets the fundamental limits of resolution and sensitivity in Dynamic Atomic
Force Microscopy (DAFM). While thermal fluctuations are conventionally assumed to be
the dominant noise source, this work demonstrates that tip--sample interactions in
ambient conditions introduce a non--thermal noise component that significantly
exceeds the thermal background. Using a model system of sodium dodecyl sulfate (SDS)
on graphite, we characterize this noise through force spectroscopy, 3D imaging
modes, and Kelvin Probe Force Microscopy (KPFM). This interaction--induced noise arises from the stochastic formation and rupture of nanoscopic liquid
necks, serving as a direct fingerprint of local wettability and dissipative
dynamics. Crucially, we find that this ``noise channel'' provides chemical contrast
that is distinct from and complementary to the electrostatic potential mapped by
KPFM. By deciphering the physical origin of these fluctuations, we establish that
noise is not merely an instrumental artifact but a rich spectroscopic signal, and we
propose that Frequency Modulation (FM--DAFM) offers a superior approach to decouple
these dissipative effects for high--resolution imaging.
\end{abstract}


\maketitle
\section{Introduction}
Nanoscale interactions arise from dispersion forces (Van der Waals and Casimir forces), short range electrostatic forces, molecular dipoles and hydrogen bonds \cite{israelachvili2011intermolecular, MohideenCasimir, GrossHydrogen, butt2010surface}. These intermolecular forces are fundamental in many fields of Biology, Chemistry and Materials Science; correspondingly a variety of experimental techniques have been developed to investigate these forces, in particular the Surface Force Apparatus \cite{TaborWinterton, israelachvili2011intermolecular} and Atomic Force Microscopy (AFM) \cite{SFM, Giessibl}. In the case of AFM, the variation of these forces is measured in a force--distance experiment \cite{Weisenhorn, butt2005force}; where the intermolecular interaction between the AFM tip (of material A) and the sample to be measured (of material B) bends the cantilever acting as force sensing element. 
Understanding and controlling in detail the tip--sample interaction is essential for AFM operation, since this technique is precisely based on this interaction, which is kept constant during imaging. Precise knowledge of tip--sample interaction is also fundamental to correctly process AFM images, and for correct data interpretation to obtain ``true topography'' \cite{HighLow} as well as correct nanoscale material characterization. Understanding the different noise sources in AFM is essential to improve both its resolution and its sensitivity. The fundamental noise sources in AFM are: thermal fluctuations of the cantilever (force sensing element) \cite{Butt}, shot noise of the laser typically used for detection, and quantum noise \cite{afmQuantum, labudaCoating}; the latter being negligible in practical applications. In addition, technical noise sources such as laser intensity fluctuations, mechanical instabilities \cite{mascaro_eliminating_2019}, cantilever coating \cite{labudaCoating} and thermal drift may degrade typical AFM experiments \cite{labuda_stochastic_2012}.

According to thermodynamics, noise in a physical system is related to fluctuations and dissipation in the system (fluctuation--dissipation theorem \cite{Nyquist1928, Callen1951, BookStatPhysics, Butt}). In this context, thermal noise in AFM arises essentially from the Brownian motion related to the whole cantilever, and thus involves many atoms. In an AFM tip--sample system, we may ask how many particles are involved in tip--sample interaction; since the system is of nanometer size we would expect a relatively small number of particles, and thus a large relative fluctuation $\Delta n/n=1/\sqrt n$, and thus a large amount of noise. Indeed, here we show that noise induced within the tip--sample nano-gap can be be significantly larger than thermal noise, which is usually assumed to be the most relevant noise source.

This work focuses on analyzing and characterizing the noise originating from tip--sample interactions in non--contact Dynamic AFM (nc--DAFM, \cite{Albrecht, Durig, Giessibl, colchero_resolution_2001, Wastl2013AmbientFM, Wastl2017QPlusAmbient, Almonte2017}), particularly in ``Amplitude Modulation'' DAFM (AM--DAFM) \cite{GarciaR, ruppert_review_2017} used in ambient conditions. Noise related to forces and tunneling occurring in the tip--sample gap of an AFM and a Scanning Tunneling Microscope due to thermal fluctuations, contact potential, non-linearity or dynamics of the cantilever motion has been the focus of a variety of investigations \cite{DuerigNoiseSTM, koslowski_new_1995, kurz_scanning_1991, labuda_stochastic_2012, chakraborty_noise_2011, howell_molecular_2002}. We will show here that using nc--DAFM in ambient conditions, the tip--sample interaction itself may introduce an additional noise component. Unlike thermal noise, which follows well--established statistical mechanics principles, this noise due to tip--sample interaction depends on material properties and specific interaction forces. Although noise is conventionally seen as a limitation, we will demonstrate how it can also serve as a diagnostic tool for nanoscale material characterization. Using techniques such as Kelvin Probe Force Microscopy (KPFM), force spectroscopy, and three--dimensional imaging modes, we aim to unravel the complex noise landscape and explore its potential for advanced AFM analysis. 
When instrumental and fundamental noise sources of a system are well controlled and known, studying noise related to this system can be used to extract fundamental statistical information about the system itself. Along these lines, we will show here that when nc--DAFM is used in humid environments, the noise induced by the tip--sample interaction is intrinsically related to the nanoscale wetting properties of the sample.

\section{Noise measurement in non--contact Dynamic Atomic Force Microscopy}

\subsection{Non--contact DAFM Data Acquisition: processing dynamic channels and time constants}
\begin{figure}[H]
\centering
\includegraphics[width=10cm]{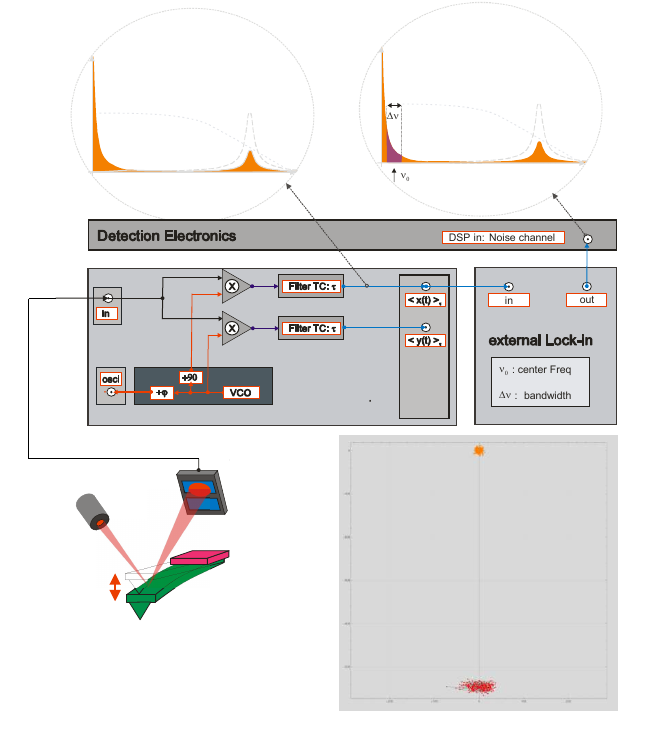}\caption{Experimental scheme for measuring noise. Essentially, the oscillation of the cantilever is excited by the signal \textit{OUT}. The signal related to the cantilever deflection enters the electronics at \textit{IN}, and is then multiplied internally with the reference signals. After the 2 quadrant detection stages, the signals are averaged over a time  $\tau_{\rm lockIn} $ giving the outputs $X_{out}(t)=<X(t)>_{\tau_{\rm lockIn}}$  and $Y_{out}(t)=<Y(t)>_{\tau_{\rm lockIn}}$.  These signals can be represented in 2D, the corresponding point allows to visualize the state of the system. At resonance, as discussed in the main text and shown in the X,Y- representation (lower right part), the output is imaginary $a(\nu_0)=i a_0 Q$ , with  $(X(t),Y(t))=(0, a_0 Q_{eff}(t) )$. If the PLL is enabled, the output  $Y_{out}(t)=0$, and is thus high -pass filtered with the lower cutoff frequency (of the high-pass) $\nu_{\rm PLL}=1/\tau_{\rm PLL}$ and the upper cut off frequency $\nu_{\rm lockIn}=1/\tau_{\rm lockIn}$ (the frequency of the low-pass). The lower cutoff frequency  $\nu_{\rm PLL}$ is determined by the PI values of the PLL feedback loop. Finally, for noise measurements the signal   $Y_{out}(t)$ is processed by an external Lock-In amplifier to measure the noise around a frequency $\nu_{\rm ref} \sim 2-4$ kHz with a bandwidth $\Delta \nu_{\rm bwNoise} \sim 0.5-1.5$ kHz}.
  \label{fig_1}
\end{figure}

To correctly measure noise, an experimental system is needed which has as little technical noise and drift as possible, such that data acquisition is essentially limited only by fundamental noise, in our case thermal noise. Here we use cantilevers with a force constant $c_{\rm lever} \sim 2-3$ N/m having a thermal noise $a_{\rm thermal} = \sqrt{kT/c} \sim 40 pm_{\rm rms}$, and thus much larger than our technical noise level $a_{\rm tech} \sim 5-10$ pm$_{\rm rms}$, which allows to detect thermal noise of much harder cantilevers (up to $c_{\rm lever} \sim 100$ N/m, see \cite{Gonzalez2013}). 
In our Optical Beam Deflection setup a key element to significantly reduce noise of our detection system has been the use of a laser system where the light is coupled into a monomode fiber to filter lateral pointing instabilities of the laser. Also, as noise measurements are quite slow (up to 11\,h) our experiments require a mechanically stable and low drift piezo scanner and AFM system. To correctly process and interpret noise data, we have used a DAFM measurement scheme similar to that discussed in more detail elsewhere for measurement and calibration of thermal noise \cite{Gonzalez2013}. As sketched in fig.~\ref{fig_1}, in our set--up we use an electronics which is essentially equivalent to a 2 quadrant Lock--In amplifier with an additional Phase Locked Loop (PLL) loop that adjusts the frequency of the signal exciting cantilever oscillation. At resonance the excitation signal and its response are $\pi / 2$ out of phase; the in--phase component (=reference signal for the PLL when enabled) of the Lock--In amplifier set--up then vanishes and the oscillation amplitude is then measured completely in the out of phase signal. As discussed in more detail elsewhere \cite{Ruido}, for high quality factor Q the state of the harmonic oscillator as a function of the driving frequency is described by a circle in the complex plane. At resonance ($\nu=\nu_0$), the oscillation amplitude is imaginary: $a(\nu_0)=i a_0 Q$ where $ a_0$ is the driving force (note however that in AFM applications a correcting factor $ k_j$ is needed due to the non--trivial properties of each cantilever mode $j$: $a({\nu_j}) =i k_j a_0 Q_{\rm eff}$ \cite{sader2019oscillation}). In this scheme, when the oscillation amplitude is kept constant by the topography feedback with the PLL enabled and the tip--sample system thus at resonance, topography feedback is performed with a constant $a_{\rm set} =i a_0 Q_{\rm eff}<a_{\rm free} =-i a_0 Q_{\rm free}$ and therefore at constant dissipation, which is described by an effective quality factor $Q_{\rm eff}$, related to dissipation within the tip--sample system. When the system is initially at resonance (cantilever ``free'', no interaction) and the PLL is either disabled, then a dissipative interaction will induce a reduction of oscillation amplitude (imaginary component$=$out of phase signal), while a conservative interaction (force gradient) will bring the tip--sample system out of resonance and induce an in--phase signal. If the system is tuned to be precisely at resonance, then conservative interactions and dissipation will induce signals that are $\pi / 2$ out of phase, which can be visualized in real time on an oscilloscope (in $xy-$mode, see fig.~\ref{fig_1}).
To correctly understand the measurement of noise in our work it is essential to understand and take into account the different time scales relevant in a DAFM experiment, and to carefully adjust these time constants as well as the feedback loops involved. The time constant $\tau_{\rm lockIn} $ of the Lock--In averaging process (fig.~\ref{fig_1}) determines up to which frequency $\nu_{\rm lockIn}=1/\tau_{\rm lockIn} $ signals detected in the normal force are ``transfered'' to the in phase and out of phase outputs of the DAFM electronics; this time constant is essentially the number of oscillations $\tau_{\rm lockIn} \nu_0$ used to ``calculate'' the dynamic outputs. The time constant $\tau_{\rm PLL} $ of the PLL--feedback response determines how fast a variation of the in--phase output is ``transferred'' to the frequency--shift channel. The time constant $\tau_{\rm topo} $ describes the ``reaction time'' of the topography--feedback response. The time constant $\tau_{\rm thermal} $ is related to thermal noise, and how fast thermal energy is exchanged with the thermal bath; $\tau_{\rm thermal} $ is essentially determined by the width of the thermal spectrum $\Delta \nu=\nu_{\rm res}/Q$. Finally, in our experiments we have a time constant $\tau_{\rm noise} $ of the noise measurement adjusted on the external Lock in amplifier (see below). In our present work, with the resonance frequency of the cantilever $\nu_{\rm res}\sim 80$ kHz we typically use: $\tau_{\rm lockIn} = 10 /\nu_{\rm res} \sim 0.125$ ms (bandwidth of $ \sim 8$ kHz, corresponding to averaging the oscillation over 10 periods), $\tau_{\rm PLL} \sim 2-5$ ms, $\tau_{\rm topo} \sim 5$ ms and $\tau_{\rm thermal} \sim 1 ms$ (corresponding to a width of the thermal spectrum $\Delta \nu=\nu_{\rm res}/Q\sim 1$ kHz, with $\nu_{\rm res}=80$ kHz and $Q_{\rm free} \sim 80$) and a large averaging time of the external Lock--in noise measurement $\tau_{\rm noise} \sim 1$ s. 
With these parameters, when AFM imaging (assume 256 pixels) is performed at 1 Hz/line, topography feedback ($\tau_{\rm topo}$) and measurement of frequency shift ($\tau_{\rm PLL}$) are performed at approximately the same rate as the acquisition time of each image pixel. Since the time constant of the Lock--In averaging process $\tau_{\rm lockIn} \sim 0.125$ ms (fig.~\ref{fig_1}) is smaller than the time constant related to thermal noise $\tau_{\rm thermal} \sim 1$ ms, our DAFM detection unit ``sees'' thermal noise in the in--phase (``phase'') and out of phase (``amplitude'') channels (for more detail see \cite{Ruido}). Moreover, even when the PLL is turned on to detect frequency shifts due to tip--sample interactions, (most of) the thermal noise still ``appears'' in the ``phase'' channel of our DAFM detection unit since $\tau_{\rm PLL} > \tau_{\rm noise}$ , the PLL is thus too slow to transfer the thermal noise detection from the ``phase channel'' to the ``frequency shift channel''. This can be clearly observed on an oscilloscope in $xy$ mode by turning on/off the PLL of our DAFM detection system, and checking that the diameter of the ``amplitude'' vs. ``phase'' output (precisely the thermal noise) does not change, which is essentially the basis for the ``dynamic'' calibration of sensitivity \cite{Gonzalez2013}. 

Generally, all possible kinds of images were acquired simultaneously; that is, during the same data acquisition session (imaging) topography, noise, normal force, oscillation amplitude, phase, frequency shift and (occasionally) also Contact Potential data (measured using Kelvin Probe Force Microscopy) are recorded. 
\subsection{Measuring noise}
To experimentally measure the noise of the tip--sample interaction, an external Lock--In amplifier was used. While in principle both, out of phase (amplitude, that is, dissipation) or in--phase (force gradient, that is, conservative interaction) components of the oscillation could be used to analyze the noise of the interaction in nc--DAFM, in the present work we have chosen the latter. On the one hand, this avoids possible issues related to coupling of the topography feedback (using the amplitude) into our noise measurements; and on the other hand, since the phase signal is kept around 0 by the PLL, this allows for high input gains of the external Lock--In amplifier used for the noise measurement. In this context we recall that the DC component of the amplitude signal is quite large (typically several volts), while the noise signal has a much smaller variation (typically only tens of mV) added to this large DC value, therefore this large DC value of the amplitude signal may easily saturate the Lock--In amplifier.
For our noise measurements the in--phase signal is therefore used as input of an external Lock--In amplifier, which is adjusted to a reference signal frequency $\nu_{\rm ref} \sim 2-4$ kHz and a bandwidth $\Delta \nu_{\rm bwNoise} \sim 0.5-1.5$ kHz. With these settings, noise of the in--phase DAFM signal (``phase'') is analyzed in a region between about 0.5 and 4 kHz, and thus within the bandwidth $\Delta \nu_{\rm bwLockIn} =1/\tau_{\rm lockIn} \sim 8$ kHz of our detection electronics. The DAFM electronics essentially ``folds'' the normal force signal to lower frequencies. More precisely: when the PLL is ``on'' the normal force signal at the resonance is brought to DC (and is essentially ``seen'' as the DC signal of the ``amplitude'' output) and the ``sidebands'' of the normal force signal around the resonance frequency are observed in the in--phase and out of phase components up to a maximum frequency corresponding to $\nu_{\rm lockIn} = 1 /\tau_{\rm lockIn} \sim 8$ kHz. With these settings, our noise measuring setup is essentially detecting noise corresponding to fluctuations of the force gradient -and thus conservative interaction- in a frequency bandwidth of 0.5--4 kHz around the resonance frequency.

\section{Results and Discussion}

\subsection{Topography, Frequency Shift and chemical inhomogeneity}

\begin{figure}[H]
\centering
\includegraphics[width=14cm]{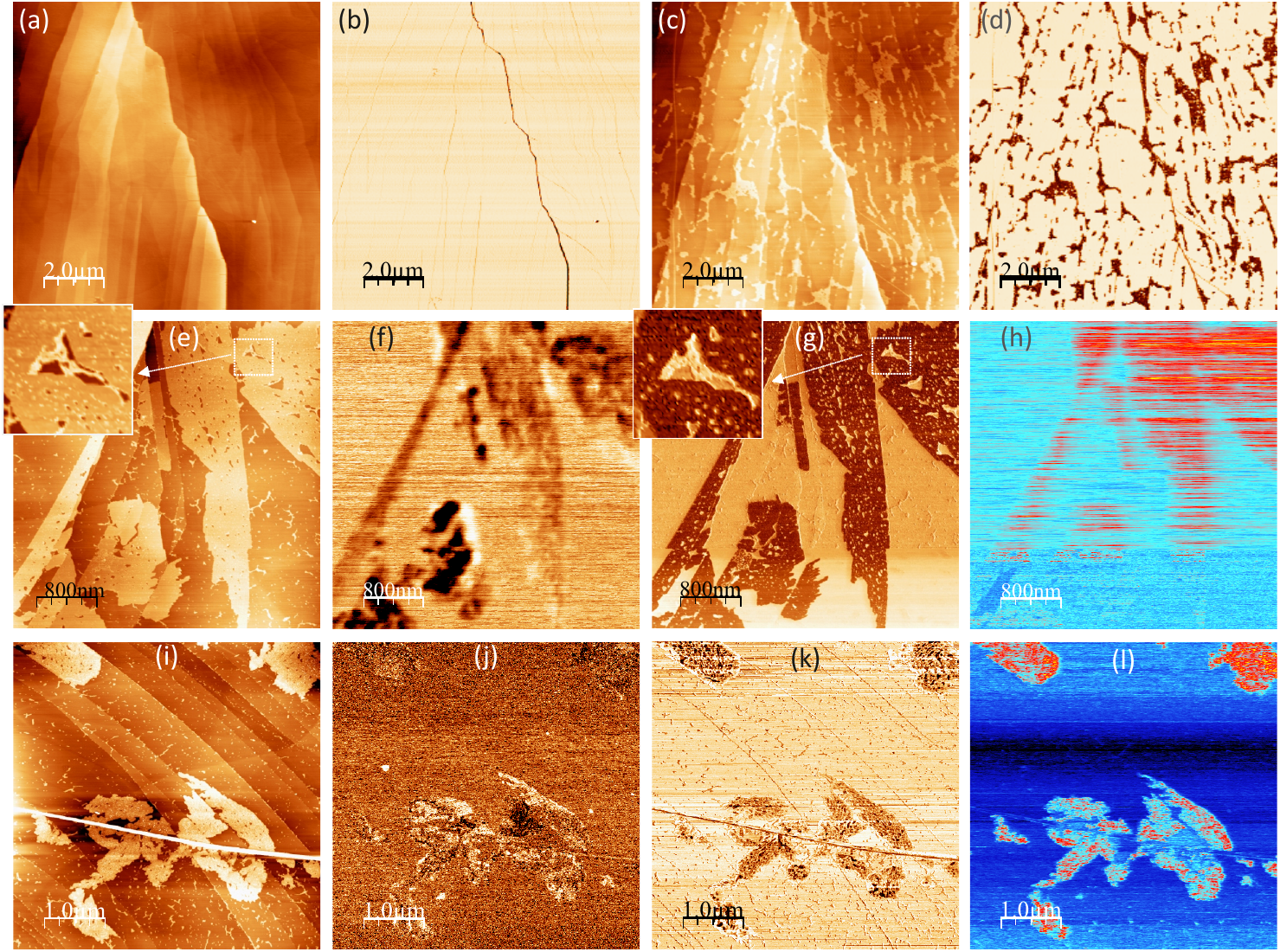}
\caption{Multidimensional characterization of the SDS/HOPG model system. 
\textbf{Top row (a--d):} Topography and frequency shift images of the same $10 \times 10~\mu$m$^2$ region before (a, b) and after (c, d) SDS deposition. Color scale ranges are $\Delta z = 40$~nm for topography and $\Delta f = 800$~Hz for frequency shift. The frequency shift clearly resolves the chemical contrast between the hydrophobic substrate and the adsorbed surfactant islands. 
\textbf{Middle (e--h) and Bottom (i--l) rows:} Simultaneous mapping of Topography (e,i), Kelvin Probe Force Microscopy (f,j), Frequency Shift (g,k), and Interaction Noise (h,l). 
The \textbf{Middle row} corresponds to a $4 \times 4~\mu$m$^2$ area with ranges: topography $\Delta z = 40$~nm, CP $\Delta V = 160$~mV, frequency $\Delta f = 40$~Hz, and noise $\Delta V = 40$~mV. 
The \textbf{Bottom row} displays a $5 \times 5~\mu$m$^2$ scan with ranges: topography $\Delta z = 4$~nm, CP $\Delta V = 200$~mV, frequency $\Delta f = 20$~Hz, and noise $\Delta V = 32$~mV. 
While the substrate signal is uniform, the SDS islands exhibit significant heterogeneity in the KPFM channel (f, j), likely due to varying molecular orientations. Insets in (e) and (g) show a zoom of a hole within the upper right SDS island with heights corresponding to mono-- and bilayers. \label{fig_2}}
\end{figure}

Choosing the appropriate surface to reveal the nature of the noise in the tip--sample interaction is not a trivial task. To avoid inconsistent data associated with either large topographical features, or finite time response and non--ideal feedback (see \cite{JuanFranPowerSpectrum2012}) we selected a sample as flat as possible, but offering a large heterogeneity of surface properties since noise is intimately related to surface chemistry. Highly oriented pyrolytic graphite (HOPG) is a well--known hydrophobic substrate \cite{LiHOPG, KozbialWetting}, atomically flat on large areas, but has nevertheless steps for easy calibration, which may also serve as anchoring sites for adsorbates. Moreover, it allows for easy and fast cleaving, in order to expose the cleanest possible surface in ambient conditions. To induce chemical inhomogeneity on the HOPG substrate, we choose the amphiphilic molecule sodium dodecyl sulfate (SDS, C\textsubscript{12}H\textsubscript{25}SO\textsubscript{4}Na). This molecule is very soluble in water forming micelles at concentrations larger than the Critical Micelle Concentration of about 8 mM \cite{MukerjeeCMC, HenzlerCMC}; which, when transferred to the graphite surface, form (sub--) monolayer films \cite{Prado2025}. The upper row of figure~\ref{fig_2} shows topography (figs.~\ref{fig_2}a,c) and frequency shift (figs.~\ref{fig_2}b,d) images of such a sample, where small islands of SDS are clearly observed. The height of these structures is around 2 nm, a typical value for this molecule \cite{Wanless1996, Prado2025}, although larger structures can also be found when multiple layers overlap (see below). Note that these images correspond to essentially the same region before (figs.~\ref{fig_2}a,b) and after (figs.~\ref{fig_2}c,d) deposition of the SDS islands on the graphite surface\cite{abad2013note}. The images were acquired with the oscillation amplitude as control parameter for the topography feedback (AM--DAFM) using a relatively small oscillation amplitude (typically $a_{\rm free} \sim 5-10$ nm) and a small amplitude reduction of 5--15 \% (that is: $a_{\rm set} \sim 0.85-0.95 a_{\rm free}$). In addition, a Phase Locked Loop (PLL) is used to keep the tip--sample system at resonance and the corresponding frequency shift $\Delta f$ of the cantilever due to tip--sample interaction is tracked by the PLL. This signal is used as secondary channel to measure differences in chemical properties of the sample. As discussed in more detail elsewhere \cite{Palacios2009}, using these parameter settings DAFM is operated in the non--contact regime, topography feedback is performed at constant dissipation, and the frequency shift $\Delta f$ is essentially proportional to the force gradient of the interaction (averaged over the oscillation distance) \cite{giessibl_forces_1997}. A brighter color in the frequency shift data shown in figs.~\ref{fig_2}b,d corresponds to a larger (attractive) interaction, that is, a more negative force gradient of the tip--sample interaction. We thus conclude that the SDS islands have a smaller attractive tip--sample interaction as compared to the graphite substrate. This is to be expected if the SDS molecules expose their hydrophobic carbon tails and the system is operated in the non--contact regime. For nc--DAFM in humid air the dissipation in the tip--sample system is due to the formation and rupture of liquid necks \cite{Men2009CapillaryLiquidBridges, Sahagun2007EnergyDissipation, marmur1993tip, Butt2009NormalCapillaryForces, colchero1998observation, Luna1999, Wei2021EnvironmentalDissipation} and the force gradient is mainly due to the strength of the liquid neck formed, which is expected to be smaller for a more hydrophobic material (see also Supporting Information in \cite{Almonte2021}).

To complement topography and frequency shift information, and to study in more detail the chemical variability of the sample we incorporated Kelvin Probe Force Microscopy (KPFM) to simultaneously map its electrostatic properties \cite{sadewasser_kelvin_2012, KPFMAmbient2025}. In addition, the noise within the tip--sample interaction was measured as discussed above. The middle and lower row of fig.~\ref{fig_2} presents the corresponding data showing topography, Kelvin (contact potential difference), frequency shift, and noise acquired simultaneously. The samples presented in the middle and lower row have a higher coverage of SDS as compared to the sample shown in the upper row; resulting in larger islands, some of them having more than a single monolayer height (see insets of figs.~\ref{fig_2}e,g). In some cases, the SDS molecules tend to align along the step edges of the HOPG substrate. 

The frequency shift, the contact potential as well as the noise images show a clear chemical contrast: for each of these properties the SDS islands generally present a different contrast as compared to the HOPG, allowing to effectively distinguish the SDS islands from the HOPG substrate. The frequency shift, the contact potential and the noise have a uniform value on the HOPG substrate. On the SDS islands, their value is much more complex and often quite nonhomogeneous. We note that the uniform value of these signals on the HOPG substrate indicates a stable tip--sample system, and in particular that no tip changes occur during imaging. 

We will first discuss the contact potential maps. On the HOPG substrate, the measured contact potential values are $V_{cp}\sim -350$ mV (fig.~\ref{fig_2}f) and $V_{cp}\sim -530$ mV (fig.~\ref{fig_2}j). This shift in the absolute baseline is attributed to a variation in the tip termination (and consequently its work function) between experiments, a well--known phenomenon in KPFM measurements \cite{Melitz2011}. The SDS structures consistently exhibit a highly non--uniform surface potential, displaying regions with both higher and lower contact potential relative to the HOPG substrate. The Kelvin channel (fig.~\ref{fig_2}f,j) thus reveals significant variations in surface potential within the islands themselves. We have not found any clear correlation between the heterogeneity of the measured contact potential and the height, frequency shift or noise data. As an exception we note, however, that for the images shown in the lower row some bright spots in the contact potential image (fig.~\ref{fig_2}j) are also clearly resolved as higher noise regions in the corresponding noise data (fig.~\ref{fig_2}l). SDS is an amphiphilic molecule that can adopt different packing arrangements (e.g. monolayers, bilayers, or hemicylinders) \cite{Prado2025}, potentially exposing either the highly polar head groups or the hydrophobic alkyl tails. The KPFM signal is sensitive to the resulting local charging, dipoles and molecular orientation \cite{Lu1999, howell_molecular_2002}. For SDS structures comprising more than a single monolayer a different orientation, thickness or stacking of the molecule may lead to quite different contact potential. Depending on the precise number of SDS layers, either the highly polar, and thus hydrophilic headgroup may be exposed, or the more hydrophobic carbon tail may be the uppermost layer of the structure. In our opinion, these different possible configurations of the molecules within the SDS islands should explain the heterogeneity of their contact potential.

For the case of the frequency shift images (figs.~\ref{fig_2}g,k), the contrast on the SDS islands is much more well defined, as compared to the case of the contact potential data. The SDS--islands present again quite consistently a smaller (attractive) interaction as compared to the HOPG. However, we note that if the islands are higher (see insets of the figs.~\ref{fig_2}e,g), these regions of the SDS islands have a much larger (attractive) interaction, even larger than the interaction measured for the HOPG substrate (see insets of figs.~\ref{fig_2}e,g). Since we operate our system in the non--contact regime, and liquid necks are an important contribution to tip--sample interaction we infer that at these locations the SDS islands are exposing their hydrophilic tails \cite{colchero1998observation}.

The frequency shift image (fig.~\ref{fig_2}k) corresponding to the sample presented in the lower row presents a similar behavior. The raw data had a large variation (see Supporting Information), presumably due to drift and tip--changes; therefore some filtering has been applied \cite{WSxM} to have the value corresponding to the HOPG at a common level. This allows to better distinguish the (relative) variation of the interaction between the HOPG substrate and the SDS--islands.
figs.~\ref{fig_2}h,l show noise images measured simultaneously with frequency shift and contact potential data. The image~\ref{fig_2}h was taken at different scan speed: 5\,s/line the upper part, and 100\,s/line the last lines at the lower part of the image. Clearly, very slow scan rates are needed in order to obtain good noise data, while essentially no difference is observed in the quality of the corresponding topographic, frequency shift and contact potential data. Both noise images present a clear chemical contrast allowing to differentiate the HOPG substrate from the SDS-island based on the amount of noise measured in the tip--sample interaction.

\subsection{Noise Imaging}

\begin{figure}[H]
    \centering
    \begin{minipage}{0.9\textwidth}
        \centering
        \includegraphics[width=\textwidth]{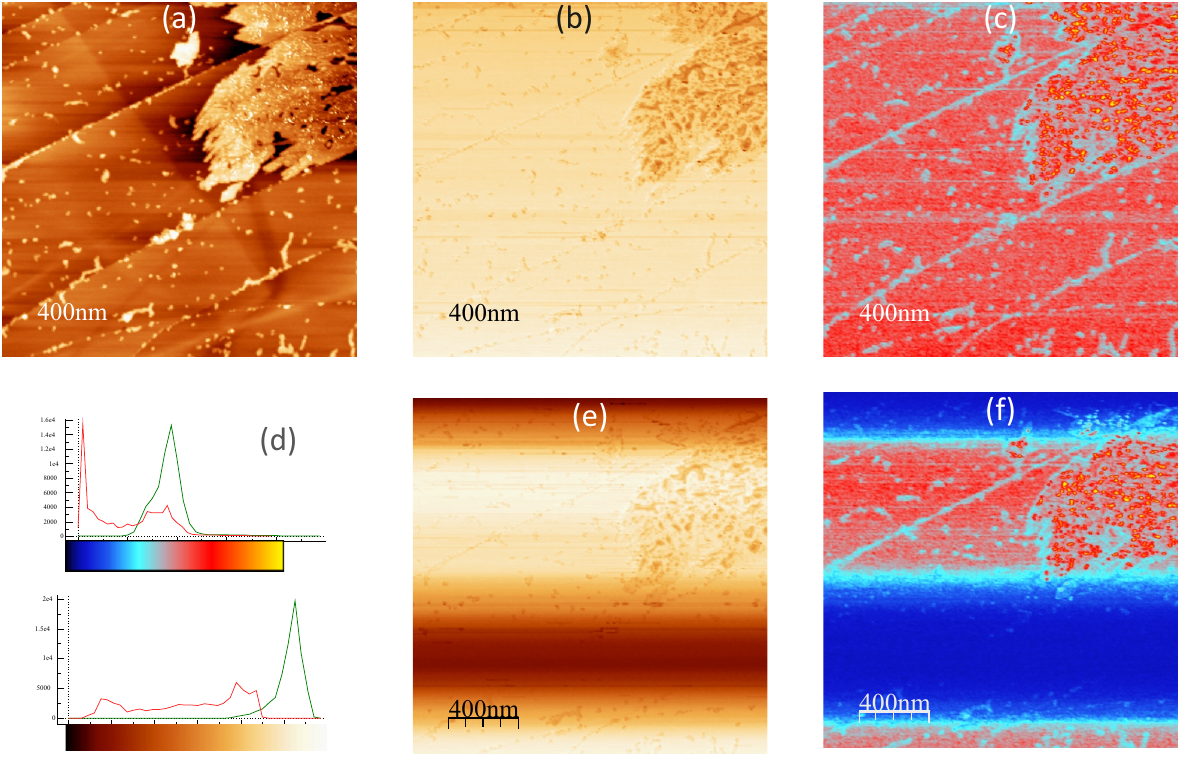}
        \caption{High--resolution mapping of interaction noise on SDS/HOPG ($2.0\,\mu\mathrm{m} \times 2.0\,\mu\mathrm{m}$).
\textbf{Top row (a--c):} Data acquired under stable interaction conditions with a grounded tip ($t_{\mathrm{acq}}\sim 5.9$~h). 
\textbf{(a)} Topography ($z$-scale: $3\,\mathrm{nm}$). 
\textbf{(b)} Frequency Shift ($z$-scale: $80\,\mathrm{Hz}$). 
\textbf{(c)} Interaction Noise ($z$-scale: $40\,\mathrm{mV}$), showing high contrast and resolution, the color scale of noise is such that low noise corresponds to (dark) blue, and high noise to red/orange, as shown also in (d).
\textbf{Bottom row (d--f):} A subsequent measurement illustrating the effect of thermal drift. Topography is omitted in this row as it is essentially identical to (a).
\textbf{(d)} Histogram analysis comparing data from stable imaging (green data from figs. b,c) and when the system is drifting (red data from figs. e,f) states. \textit{Top:} Noise histograms; green curve corresponds to (c), red curve to (f). \textit{Bottom:} Frequency histograms; green curve corresponds to (b), red curve to (e).
\textbf{(e)} Frequency Shift ($z$-scale: $80\,\mathrm{Hz}$).
\textbf{(f)} Interaction Noise ($z$-scale: $40\,\mathrm{mV}$). Its color scale is the same as for the noise image (c), and shown with the Histogram analysis in (d) .\label{fig_3}}
    \end{minipage}
\end{figure}

Figure~\ref{fig_3} shows topography, frequency shift and noise data measured simultaneously on a SDS covered HOPG surface. No ac--voltage bias is applied to the tip in this case and the tip is grounded to avoid possible degradation of the noise signal due to coupling of the electrical excitation needed for contact potential measurements. As in the previous images the HOPG substrate is clearly distinguished from the SDS islands. We find small SDS islands with a size of the order of 20$-$50 nm and a well defined height, and a much larger structure (almost 1$~\mu$m) at the upper right corner similar to the larger structures covering the HOPG structure in figure~\ref{fig_2}. Again, this larger structure has a non--uniform height with height of different layers, and the corresponding frequency shift image on this larger structure is quite heterogeneous. We note that all images present a very high resolution, of the order of 10$-$20 nm, which is expected for nc--DAFM for a small tip radius ($R_{\rm {tip}} \sim 25$ nm) at an effective tip--sample distance $d_{\rm {eff}}$ of 5$-$10 nm using the simple, but quite good, approximation: resolution $ \sim \sqrt{R_{\rm{tip}}d_{\rm{eff}}}$ \cite{Palacios2009, colchero_resolution_2001}.

The noise images also have a very high resolution, if acquired with very large acquisition time: the scan rate is about 80\,s/line (6\,h total time). Feedback is then essentially ``perfect'' since the topography feedback loop has sufficient time to adjust at each image point (see supporting information), therefore excluding issues related to finite feedback response, in particular at step edges. The images were taken with the smallest interaction possible, with amplitude reduction of 5$-$10\% of the free oscillation frequency. The color scale of the noise data is such that low noise corresponds to (dark) blue, and high noise to red/orange. Due to the large acquisition time, a small amount of thermal drift may slightly change the free oscillation amplitude $a_{\rm set}$, resulting in a variation of the effective set-point $a_{\rm set}/a_{\rm free}$, and thus an effective variation of interaction. This is observed in the lower row, where the effective variation of amplitude set--point results in the dark regions in the noise image: as the true interaction at which image is acquired varies slightly, the tip--sample distance varies. For the smallest effective tip--sample interaction the noise is low (fig.~\ref{fig_2}f); when the interaction is larger (brighter region in the $\Delta f $ image,~fig.~\ref{fig_2}e), the noise increases significantly (red regions in the noise image,~fig.~\ref{fig_2}f).

We note that the variation of interaction is also observed in the frequency shift and the topographic images: where the interaction is lowest (dark band in the lower part of the $\Delta f $ image, fig.~\ref{fig_2}e) the frequency shift image loses contrast and the features in the topographic images look ``less sharp'', corresponding to a lower resolution. This is precisely what is expected when imaging at low interaction, and thus at a larger tip--sample distance $d_{\rm {eff}} $. We thus conclude that the interaction (amplitude $=$ ``dissipative'' as well as frequency shift $=$ ``conservative'') and also the noise varies with tip--sample distance, which explains the variation of frequency shift as well as noise while the image is acquired (see next section, where tip--sample interaction is discussed in detail). 

While the variation of set--point during image acquisition is of course non--ideal, it also allows to study how the image varies with different set--points. In this respect, we find the noise image quite intriguing. In particular: its contrast changes with interaction strength when comparing the small islands compared to the HOPG substrate. For small interaction (darker parts at top of the frequency image, fig.~\ref{fig_2}e, and lower dark band) the noise on the islands (fig.~\ref{fig_2}f) is larger than on the HOPG substrate, while for larger interaction (brighter ), the islands show a lower noise as compared to the HOPG substrate. Note that the contrast of the $\Delta f $ image does not vary for the different effective interactions sampled in this image: the small SDS islands always have a smaller (attractive) tip--sample interactions as compared to the HOPG substrate (excluding, of course, the region were the contrast is lost completely). We thus conclude that frequency shift and noise images are not directly related; that is, the noise image is not simply a ``copy'' of the frequency shift image. This conclusion is also supported by comparing frequency shift and noise measured in the region corresponding to the larger structure at the upper right corner: clearly both images are quite different in structure and resolution. Interestingly the noise images, when taken sufficiently slow, have a higher resolution and look ``cleaner''.

\subsection{Interaction and noise as a function of distance: noise spectroscopy}

To further investigate the physical origin of the interaction noise, we performed force spectroscopy experiments \cite{Weisenhorn,butt2005force} at specific locations on the sample: for a fixed lateral position the normal force, the oscillation amplitude and the noise data where acquired as a function of tip--sample distance. This technique allows to decouple the distance--dependent behavior of the interaction and the noise from lateral variations of theses signals due to the different material properties of the sample.

\begin{figure}[H]
    \centering
    \begin{minipage}{0.9\textwidth}
        \centering
        \includegraphics[width=\textwidth]{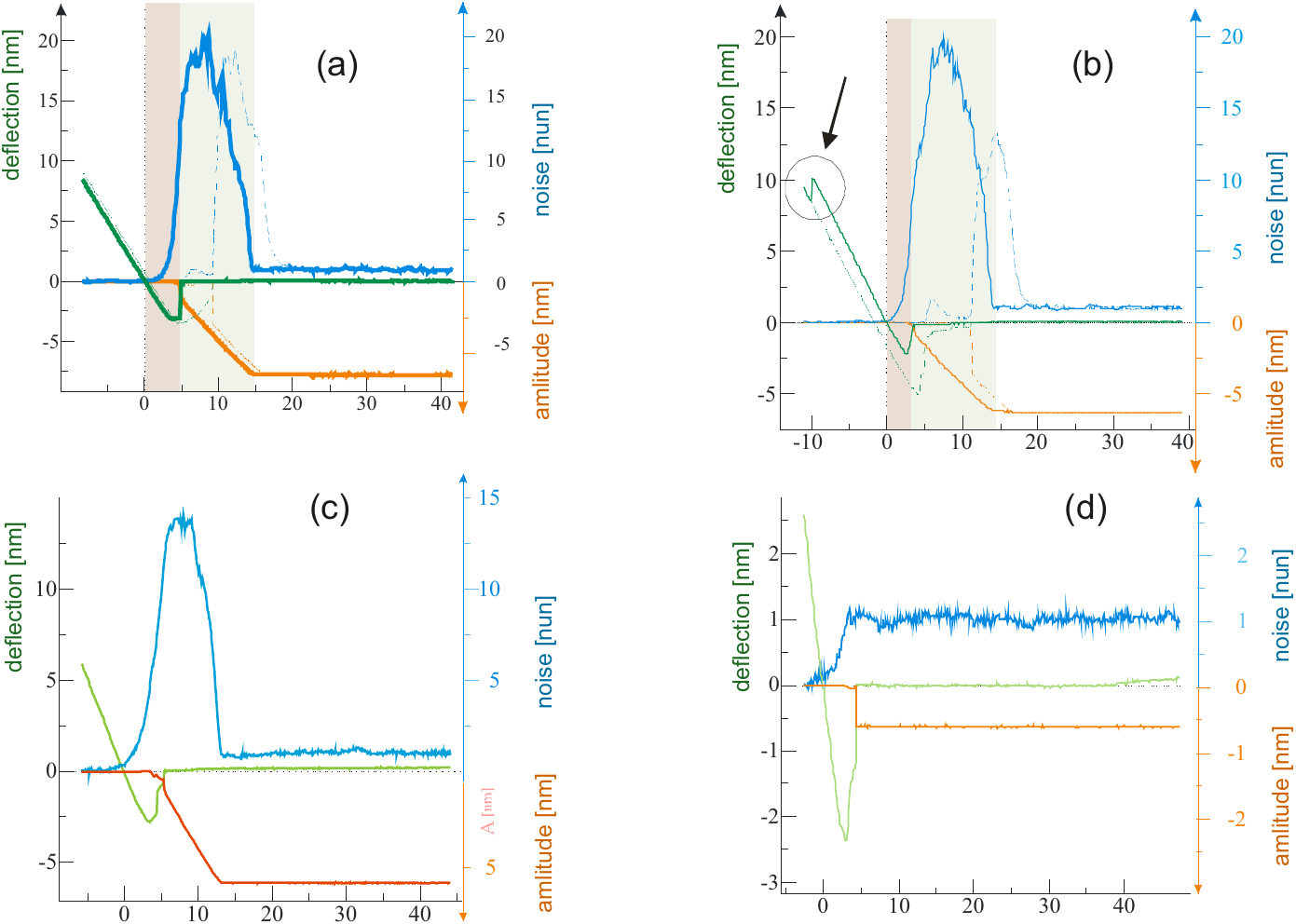}
        \caption{Spectroscopic analysis of interaction noise on SDS and HOPG. Simultaneous curves of Normal Force (green), Oscillation Amplitude (orange), and Noise (blue) versus tip--sample distance $\Delta$. \textbf{Top row (a, b):} Measurements with standard oscillation amplitudes ($a_{\rm free} \sim 10$~nm). (a) Curves on the HOPG substrate showing. (b) Curves on an SDS island showing mechanical indentation (arrow). In both curves a sharp rise in noise (blue) -up to 20 times the thermal floor (NUN) - is observed as the oscillation amplitude (orange) decreases. \textbf{Bottom row (c, d):} Control experiments performed using large (c) and small (d) oscillation amplitudes ($a_{\rm free} < 1$~nm) at the same location. For samll oscillation, the noise signal remains at the thermal background level (approx. 1 NUN) throughout the whole approach range. \label{fig_4}}

    \end{minipage}
\end{figure}

Noise data were normalized to the noise level at large tip--sample distance, where the interaction is negligible and only ``thermal noise'' is measured. With the different time constants and parameters used in our experiments, most thermal noise should be detected by the external Lock--In measuring noise, therefore 1 normalised unit of noise (NUN) corresponds to the thermal noise floor at large tip--sample distances. Normal force and oscillation amplitude are measured both as ``cantilever deflection'', and thus using nm, for easier comparison of both signals. The results presented in the upper row of fig.~\ref{fig_4}correspond to spectroscopy experiments at different locations of the sample corresponding to the HOPG substrate (fig.~\ref{fig_4}a), and the SDS island (fig.~\ref{fig_4}b) respectively. The same experimental parameters (speed, oscillation amplitude and driving frequency, data points) during the spectroscopy experiments where used for the data shown in the upper row; while the lower row (fig.~\ref{fig_4}c and d) shows data corresponding to the same location, but using different oscillation amplitude without changing other parameters. To facilitate the interpretation of data, the different interaction ranges have been shaded: darker for regions corresponding to mechanical contact after the ``jump to contact'' instability, lighter for regions corresponding to the decrease of oscillation amplitude, and no shading for regions where the cantilever is ``free'' and tip--sample interaction is negligible. 

The force vs distance curves (FvsD) in the upper row show that the jump to contact distance is larger on the HOPG as compared to the SDS (5 nm vs 4 nm). Interestingly, for the FvsD curve measured on the SDS a sudden variation of approx. 2 nm is measured, corresponding to the indentation through the SDS layer (indentation/force $\sim 10$ nm/$20$ nN, see arrow). Upon retraction, a clear difference of the FvsD curves on HOPG and SDS is also observed: while the retraction FvsD curve on HOPG shows a linear reduction region (negative slope) before full detachment, the curve on the SDS has a larger adhesion (possibly due to the indentation) and detaches more suddenly. 

For large oscillation amplitude, all noise versus distance curves show a large increase of noise when the oscillation amplitude decreases but still essentially no force is detected. In this region the tip--sample system in thus in the non--contact regime. The noise increases significantly as the tip--sample interaction increases, and the oscillation amplitude decreases. The normalised noise level has typically a peak slightly before the snap to contact instability occurs, reaching values up to 20~NUN. No significant difference is observed for noise versus distance curves measured on HOPG as compared to SDS--islands. The measurement noise reveals a distinct hysteresis: noise levels are generally higher during the approach phase than during retraction (however: since noise measurements are very slow, some hysteresis may be due to the large time constant $\tau_{\rm noise}$). The noise images shown in this work are acquired at low interactions, and thus at the initial stages of the noise versus distance curves shown here, up to noise values of 5 NUN, and thus approximately 5 times thermal noise. 

A significant difference is observed for noise versus distance curves measured with small oscillation amplitudes ($a_{\rm free} <1$ nm): no increase of noise is measured, the noise is essentially either thermal noise when the cantilever is far from the sample; or no noise is measured when the cantilever is in mechanical contact. 

From this we may draw two in our opinion fundamental conclusions. First, noise is intimately related to reduction of oscillation amplitude and thus to the Amplitude Modulation DAFM mode. If experiments are performed in ambient conditions using the Frequency Modulation DAFM mode \cite{Palacios2009, Wastl2013AmbientFM, Wastl2017QPlusAmbient, Santos2021HydrationDynamics}, only thermal noise is measured, no additional noise induced by tip--sample interaction. Second, our measurements clearly suggest that the noise is intrinsically linked to dissipative processes occurring at the nanoscale, specifically within the range where attractive forces dominate. 

\subsection{Spatial Mapping of Noise: 3D (x,y) Modes}

\begin{figure}[H]
    \centering
    \begin{minipage}{0.9\textwidth}
        \centering
        \includegraphics[width=\textwidth]{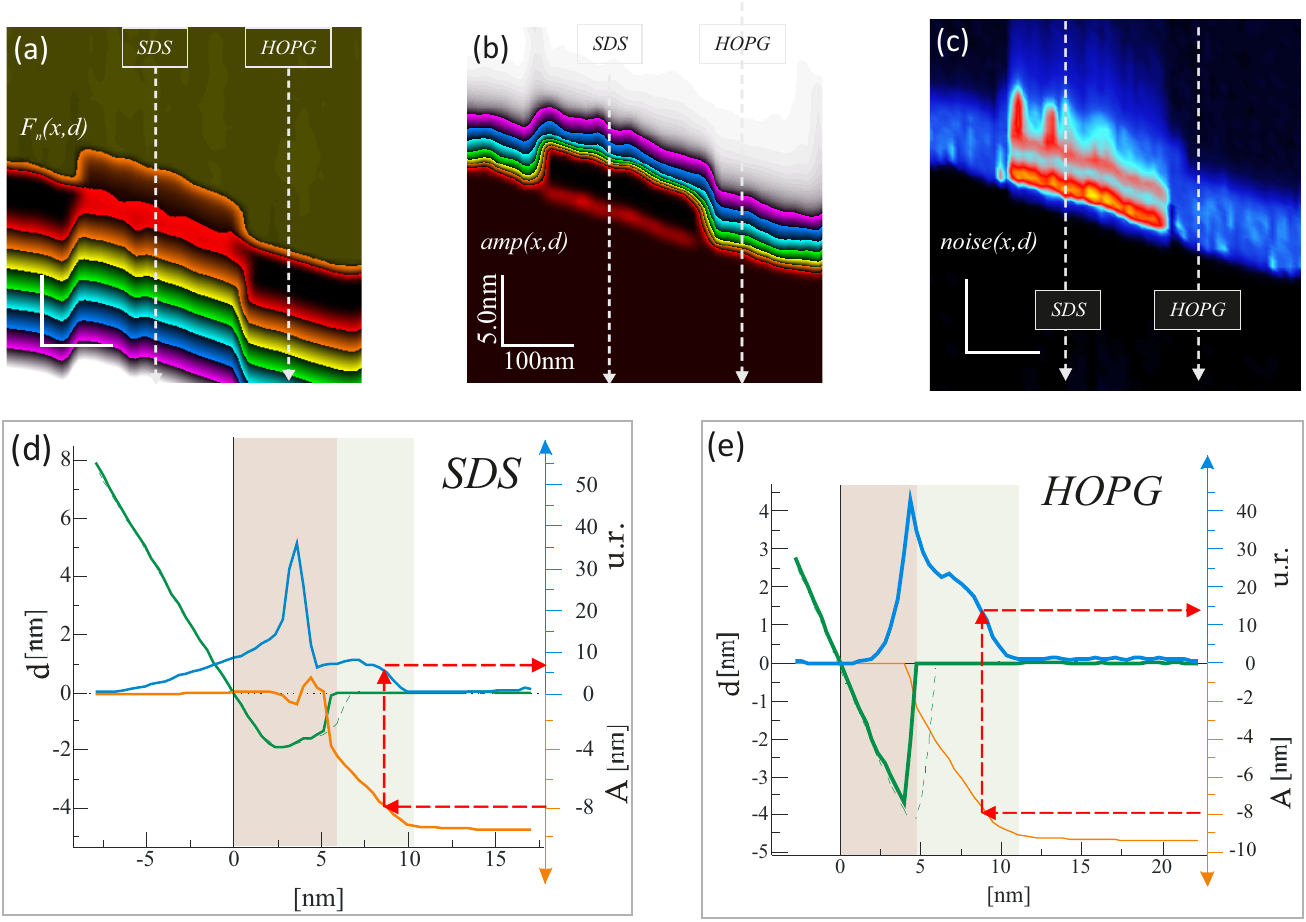}
        \caption{Spatial dissection of interaction dynamics: 3D $(x,d)$ analysis. Maps of \textbf{(a)} normal force $f_n (x,d)$, \textbf{(b)} amplitude $amp (x,d)$, and \textbf{(c)} Interaction Noise $noise (x,d)$ acquired by stacking spectroscopy curves along a scan line crossing an SDS island during an approach cycle, that is from non--contact to contact. The horizontal $x$ axis represents the lateral position (slow scan) and vertical $d$ axis represents the tip--sample distance (fast scan). While the amplitude data (b) displays a gradual onset, the noise map (c) shows a much more complex behavior. \textbf{(d,e)} Normal force $f_n (d)$, oscillation amplitude $amp (d)$, and noise $noise (d)$ vs. distance curves averaged for locations corresponding to the SDS islands (d) and the HOPG substrate. These data sets follow the same structure as those shown in fig.~\ref{fig_4} ; in particular: normal force is shown green, oscillation amplitude orange, and noise blue; and the different interaction ranges have been shaded darker for regions corresponding to mechanical contact after the ``jump to contact'' instability, lighter for regions corresponding to the decrease of oscillation amplitude, and no shading for regions where the cantilever is ``free'' and tip--sample interaction is negligible. \label{fig_5}}
    \end{minipage}
\end{figure}

To bridge the gap between single--point spectroscopy and full--surface imaging, we employed 3D $Interaction(x,d)$ scanning modes, measuring how normal force, oscillation amplitude and noise data vary along a lateral scan, that is, by recording spectroscopy data along a whole line scan: normal force $f_n (x,d)$, amplitude $amp(x,d)$ and noise $noise(x,d)$ are acquired simultaneously as tip--sample distance ($d$, fast scan) and lateral position ($x$, slow scan) is varied. fig.~\ref{fig_5} presents the result of a typical experiment, which provides a comprehensive view of how the interaction noise is distributed laterally and vertically. In these images, large tip--sample distances correspond to the upper part, where essentially no interaction and no noise is measured. The lower part of these images corresponds to the region of tip--sample contact (note: the normal force increases linearly in this regime). A false color scaling of amplitude and normal force is utilized that allows to infer topography data from these data sets \cite{HighLow, Almonte2021}: essentially, a topography scan can be calculated from these $Interaction(x,d)$ data sets by solving the implicit equation $Interaction(x,z(x))=I_{set}$ for the tip--sample distance $z(x)$. This can be done graphically by just following a fixed color value in the normal force $f_n (x,d)$ data set (fig.~\ref{fig_5}a), if images are in contact at a specified $f_{\rm set}$; or -as in our case- in the amplitude $amp(x,d)$ data sets, if images are in (non--contact) Amplitude Modulation DAFM mode at a specified $a_{\rm set}$. Interestingly, the height of the SDS is about 2.2 nm when determined following the equipotential lines in the $f_n (x,d)$ data, and only 1.6 nm when following the equipotential lines in the $amp(x,d)$ data. As discussed in more detail in \cite{HighLow, Almonte2021}, this can be explained by differences in the force and amplitude vs distance behavior due to the different chemistry of the SDS islands and the HOPG substrate. From the $noise(x,d)$ data we conclude that significantly higher noise occurs over the SDS regions as compared to the bare HOPG, even though the oscillation amplitude decreases in a very similar way on both regions. Fixing the lateral position $x_0$ in these data sets $f_n (x_0,d)$, $amp (x_0,d)$, and $noise (x_0,d)$; and averaging over several $x$ positions (corresponding to the same material) we obtain normal force $f_n (d)$, oscillation amplitude $amp (d)$, and noise $noise (d)$ vs. distance curves. figs.~\ref{fig_5}d,f show these averaged curves for the SDS islands and the HOPG substrate. Essentially, these curves are very similar to the curves obtained above on SDS and HOPG and discussed in fig.~\ref{fig_4}; confirming our main conclusion: noise arising from the tip--sample interaction is observed when the oscillation amplitude decreases due to dissipation within the tip--sample system.

\section{Conclusion}

By systematically characterizing a model system of SDS on HOPG, we identified an interaction--induced noise that may exceed the thermal background noise by more than an order of magnitude. Our data -images, spectroscopic data and 3D imaging analysis- show that this additional noise is intrinsically related to the reduction of amplitude, and thus to dissipation occurring in the tip--sample system. At low amplitude reduction setpoints, that is, at low dissipation, this noise is lower. No interaction--induced noise is measured when DAFM is operated in FM--DAFM mode.

In ambient conditions, on the one hand, a thin film of water will condense on a surface, whose thickness is $d=( A / (6 \pi n k T \ln(\mathrm{rh})))^{1/3}$, with $A$ Hamaker Constant, $k T$ the thermal energy, $rh$ the relative humidity and $n$ the number density of water. On the other hand, a liquid meniscus may be stable in nano--sized gaps due to the so--called Laplace Pressure $\Delta p = \gamma/\kappa = n k T \ln(\mathrm{rh})$, with $\gamma$ the surface energy of water and $\kappa$ the Kelvin Radius \cite{israelachvili2011intermolecular,YangNature,UhligNanoLett}.

Dissipation of energy in DAFM applications has been attributed to the formation and rupture of liquid necks \cite{colchero1998observation,Luna1999,Zitzler2002CapillaryForcesTappingMode,Sahagun2007EnergyDissipation,UhligNanoLett,Wei2021EnvironmentalDissipation}, which leads to hysteresis in the force vs distance curves, and thus to dissipation (in fact, this dissipation is the area enclosed by the F--d curve in each cycle). The liquid bridge may form either after contact of the liquid films when the tip--sample distance is smaller than the sum $d_{\mathrm{tip}} + d_{\mathrm{sample}}$ of the liquid films on tip and surface \cite{Zitzler2002CapillaryForcesTappingMode, Santos2021HydrationDynamics}, or due to the spontaneous formation of a liquid neck when the tip--sample distance is smaller than about twice the Kelvin radius \cite{Luna1999,colchero1998observation,Sahagun2007EnergyDissipation}. For medium and higher humidity previous work rather supports the latter case; and for a large tip--sample radius one finds that a liquid bridge should condense for a critical tip--sample distance smaller than $d_{\mathrm{liqNeck}} = \kappa (\cos(\vartheta_{\mathrm{sample}}) + \cos(\vartheta_{\mathrm{tip}}))$, where $\vartheta_{\mathrm{sample}}$ and $\vartheta_{\mathrm{tip}}$ are the contact angles of water on the sample and on the tip, respectively. Note that in this latter case, liquid neck formation depends on the wettability of the sample \cite{Almonte2021} , and thus on its chemical properties;precisely as observed in the present work.

Collectively, our observations support the hypothesis that the additional noise is generated by the stochastic formation and rupture of liquid necks between the tip and the sample during each oscillation cycle, a chaotic process that introduces fluctuations in the force gradient detected as interaction noise. In this context, we note that dissipation due to the rupture of liquid necks has been proposed to induce sub--harmonics and sidebands in the oscillation of the cantilever \cite{Sahagun2012AdhesionHysteresisPRB, SantosSubharm, Men2009CapillaryLiquidBridges, ChiesaSubharm}. We believe that our data are compatible with this interpretation; however, due to the small number of particles involved in the formation of liquid necks, instead of observing well--defined sub-harmonics of the cantilever oscillation, we observe a whole range of sub--harmonics due to fluctuations and the stochastic nature of the liquid neck. Spontaneous condensation of water molecules within the gap is a first--order phase transition involving nucleation, where critical fluctuations may become relevant. In particular, condensation of water may occur sooner or later during each tip--sample approach of the oscillation and the liquid neck may also break sooner or later as the tip retracts, leading to noise rather than in well--defined sidebands of cantilever oscillation.

Our study clearly challenges the conventional view that thermal fluctuations constitute the ultimate resolution limit in ambient Dynamic Atomic Force Microscopy (DAFM). Moreover our findings have fundamental implications for the operation of AFM in humid environments: Amplitude Modulation (AM--DAFM) is intrinsically limited by this interaction noise, as the amplitude feedback loop is directly susceptible to these stochastic fluctuations near resonance. Conversely, Frequency Modulation DAFM emerges as a superior strategy for high-resolution, low-noise imaging \cite{Albrecht,Palacios2009, Wastl2013AmbientFM, Wastl2017QPlusAmbient}, allowing for the separation of conservative frequency shifts from the dissipative noise floor. Beyond identifying the physical origin of these fluctuations, our results establish interaction noise as a distinct spectroscopic observable, independent of electrostatic measurements. While Kelvin Probe Force Microscopy (KPFM) effectively maps the static electrostatic landscape -revealing molecular dipole orientations and surface potential- it remains insensitive to the stochastic wetting processes that generate noise. Consequently, the combination of noise mapping and KPFM provides a method to decouple capillary-induced dissipation from electrostatic forces, offering a multidimensional characterization of the surface that neither technique can achieve alone.

Finally, from a much more general point of view, the tip--sample system may be considered a model system for studying the behavior of nano--sized gaps and cavities. We thus propose the AFM tip--sample system in combination with the methodology for noise measurements as a very sensitive and high--resolution system to investigate the behavior of nano--confined liquid systems \cite{YangNature, Bocquet1998MoistureInducedAgeingNature, BocquetCharlaix2010, marmur1993tip, Evans1986FluidsNarrowPoresJCP} where its small size (Kelvin radius compared to tip radius) and the small number of molecules may lead to very different behavior as compared to larger systems. In particular for the case of water, we expect water confined in nano gaps to be very different as compared to ``bulk'' water \cite{FumagalliLowDielectricScience2018}, since water has -due to hydrogen bonding network- a quite large interaction range ($\sim 2$ nm, \cite{israelachvili2011intermolecular}). Our work could thus contribute to further investigate the unique and complex properties of water \cite{Pettersson2016WaterMostAnomalousChemRev}. 

\section*{References}
\bibliographystyle{iopart-num} 
\bibliography{References_revised_final} 

\end{document}